\documentclass[12pt, letterpaper]{article}
\usepackage{cite}
\usepackage{amsmath,amssymb,amsfonts}
\usepackage{algorithmic}
\usepackage{graphicx}
\usepackage{textcomp}
\usepackage[ruled,vlined]{algorithm2e}
\def\BibTeX{{\rm B\kern-.05em{\sc i\kern-.025em b}\kern-.08em
    T\kern-.1667em\lower.7ex\hbox{E}\kern-.125emX}}
\begin{document}
\LinesNumbered

\title{Distributed-Memory Load Balancing with Cyclic Token-based Work-Stealing Applied to Reverse Time Migration}
\author{\'{I}talo A. S. Assis, Ant\^{o}nio D. S. Oliveira, Tiago Barros,\\ Idalmis M. Sardina, Calebe P. Bianchini,\\ Samuel Xavier-de-Souza}

\maketitle

\begin{abstract}
    %Contextualization
    Reverse time migration (RTM) is a prominent technique in seismic imaging. Its resulting subsurface images are used in the industry to investigate with higher confidence the existence and the conditions of oil and gas reservoirs. Because of its high computational cost, RTM must make use of parallel computers.
    %Gap
    Balancing the workload distribution of an RTM is a growing challenge in distributed computing systems. The competition for shared resources and the differently-sized tasks of the RTM are some of the possible sources of load imbalance. Although many load balancing techniques exist, scaling up for large problems and large systems remains a challenge because synchronization overhead also scales.
    %Purpose
    This paper proposes a cyclic token-based work-stealing (CTWS) algorithm for distributed memory systems applied to RTM. The novel cyclic token approach reduces the number of failed steals, avoids communication overhead, and simplifies the victim selection and the termination strategy.
    %Methodology
    The proposed method is implemented as a C library using the one-sided communication feature of the message passing interface (MPI) standard.
    %Results and discussion
    Results obtained by applying the proposed technique to balance the workload of a 3D RTM system present a factor of $14.1\,\%$ speedup and reductions of the load imbalance of $78.4\,\%$ when compared to the conventional static distribution.

\textbf{keywords:} Load balancing, Reverse Time Migration, Work-Stealing, One-Sided Communication, Distributed Memory.
\end{abstract}

%\titlepgskip=-15pt

\section{Introduction}
\label{sec:introduction}
    % Contextualization
    The migration of seismic data is the process that attempts to build an image of the Earth's interior from recorded field data. Migration places these data into their actual geological position in the subsurface using numerical approximations of either wave-theoretical or ray-theoretical approaches to simulate the propagation of seismic waves \cite{Yilmaz2001a}.
    
    The wave-theoretical approach to the propagation of seismic waves employs the finite difference method (FDM) \cite{Alterman1968,Kelly1976} to numerically solve the equation describing the movement of the waves \cite{Yilmaz2001a,Claerbout1972}. This approach is prevalent among the geophysical community, due to its capacity of dealing with substantial velocity variations in complex geology (e.g., pre-salt).
    
    Reverse time migration (RTM)~\cite{Hemon1978,Baysal1983,McMECHAN1983,Whitmore1983,Kosloff1983} implements this approach. It is one of the most known FDM-based migration methods. RTM is computationally intensive in terms of data storage and handling, and its use of high-complexity algorithms. Therefore, exploiting parallelism is mandatory for RTM implementations in 3D Earth models (3D RTM) \cite{Araya-Polo2009}.
    
    Parallel architectures can be classified as shared memory, when there is a single memory address space available to all processing units (e.g., nodes or cores), or distributed memory otherwise \cite{Diaz2012}. Many scientific and industrial computational resources are distributed memory systems composed of multi-processor nodes with shared memory systems. A hybrid parallel application works at these two levels of parallelism. It can distribute the total workload among the nodes of a distributed memory system. Each node, then, distributes its subset of the workload among the processing units of its shared memory system. Parallel machines can also be described as heterogeneous when they have processing units built from different types of hardware, or homogeneous otherwise \cite{Pacheco2011}.
    
    One of the main concerns in parallel computing is the efficient use of the available computational resources. Some applications such as RTM may suffer from load imbalance. A way of dealing with this issue is to employ load balancing techniques, which usually refer to the distribution of the workload among the available computational resources (e.g., nodes, processors, cores). The objective is to minimize the idling of the computational resources while there are still tasks remaining to be processed.
    
    % State the Gap
    Ensuring the load balancing for an RTM is especially challenging in distributed memory systems. Distributing the workload in equal amounts of tasks for each computational node may not be optimal. Even for homogeneous computational systems with an evenly distributed workload, several factors may be a source of load imbalance. It can be intrinsic to the application itself or caused by program-external factors such as runtime environment routines (e.g., system calls) and resource availability. The competition for shared resources, such as the parallel file system or the network, can cause idling due to resource contention as the availability of the resources may differ across the nodes and along time.
    
    Work-stealing (WS) is one of the main load balancing strategies. The fundamental idea of WS methods is that idle processing units steal tasks from the others \cite{Blumofe1999} in an attempt to avoid the performance overhead of a centralized entity being responsible for the task scheduling. Processes stealing tasks are the thief processes, whereas the processes with stolen tasks are the victim processes. Nevertheless, in the context of distributed systems, some WS implementations present problems with too many failed steal attempts, with communication overhead, with the victim selection, and with the termination strategy.
    
    % Show the State-of-the-Art: next section
    % State the purpose of the paper
    This paper proposes a cyclic token-based work-stealing (CTWS) algorithm for distributed memory systems applied to RTM. The novel cyclic token approach reduces the number of failed steals, avoids communication overhead, and simplifies the victim selection and the termination process. The proposed work-stealing method was implemented in C using the message passing interface (MPI)~\cite{MPI1994} standard. The communication was implemented by remote memory access (RMA) using MPI one-sided communication \cite{Gropp1999}. This communication model allows the thief processes to perform the work-stealing without directly involving (or interrupting) the victim processes, thus further reducing communication overhead. Our 3D RTM code was implemented in C using MPI for distributed-memory parallelism across nodes, and OpenMP \cite{OpemMP} for thread-level parallelism within nodes.
    
    The contribution of this paper to the fields of distributed load balancing and RTM are:
    \begin{enumerate}
    \item the proposition of a novel approach to implementing load balancing with WS in distributed systems based on a cyclic token;
    \item the mitigation of important WS implementation problems in distributed systems by the proposed cyclic token approach as it avoids failed steals and simplifies the victim selection and the termination strategy;
    \item the reduction of the communication overhead of the WS distributed implementation by the use of MPI one-sided communication to implement the cyclic token approach;
    \item a detailed evaluation of the conventional load balancing technique for 3D RTM showing that load imbalance is significant due to resource contention;
    \item improvements in the execution time of 3D RTM of about $14\,\%$ and reductions of the load imbalance of about $78\,\%$.
    \end{enumerate}
    
    The rest of this paper is organized as follows. Section \ref{sec:rtm} shows the basics of RTM and describes our RTM implementation. Section \ref{sec:ctws} introduces the work-stealing method proposed in this work. Section \ref{sec:wsrtm} details the application of the proposed technique to the RTM. Section \ref{sec:results} discusses the performance of the RTM with and without the proposed approach. Section \ref{sec:literature}  presents a literature review of related works contrasting them with the proposed approach. Finally, Section \ref{sec:conclusions} summarizes this work and proposes future research.

\section{RTM and Static Load Balancing}
\label{sec:rtm}
    In a seismic reflection survey, an acoustic source at a given location (a ``seismic shot'') generates a wave that propagates into the subsurface. Each time the wave travels through an interface between two layers with different impedance, part of its energy is reflected and is eventually registered at a set of receivers. This procedure is repeated for different shot locations in order to cover the whole area of interest. The data recorded by a single receiver for a single seismic shot is called a seismic trace, and a set of traces is called a seismogram. The seismograms can pass through many processing steps to finally provide an image of the subsurface.
    
    Migration is one of the most critical steps in processing seismic data. It aims to position the reflection interfaces properly in the subsurface. A migrated section is an image representing the geological structures in the region of interest. This section can be used for interpretation purposes, often to locate and characterize oil and gas reservoirs.
    
    Reverse time migration (RTM) \cite{Baysal1983,Kosloff1983} is one of the most known migration methods. The main steps of an RTM are presented in Algorithm \ref{alg:serialrtm}. The first step, the forward propagation, simulates the incident wavefield by propagating a source wavelet through the region of interest. The backward propagation generates the reflected wavefield by propagating the seismogram comprised of the seismic traces from a shot, a common shot gather, in reverse time order.
    
    \begin{algorithm}
    \caption{Main steps of a reverse time migration}
    \begin{algorithmic}[1]
    \label{alg:serialrtm}
    \FORALL{(shots locations)}
        \STATE forward propagation
        \STATE backward propagation of the common shot gather
        \STATE image condition
    \ENDFOR
    \end{algorithmic} 
    \end{algorithm}
    
    Both forward and backward propagation can be performed by iteratively solving, over a discrete grid, the acoustic wave equation, described as: 
    \begin{equation} \label{eq:ondaRTM}
    \frac{\partial^2 u(\mathbf{x})}{\partial x_1^2} + \frac{\partial^2 u(\mathbf{x})}{\partial x_2^2} + \frac{\partial^2 u(\mathbf{x})}{\partial x_3^2} = \frac{1}{c(\mathbf{x})^2}\frac{\partial^2 u(\mathbf{x})}{\partial t^2} + s(t).
    \end{equation}
    In \eqref{eq:ondaRTM}, $\mathbf{x} = (x_1,x_2,x_3)$ are the spatial dimensions, $u(\mathbf{x})$ is the pressure wavefield, $c(\mathbf{x})$ is a velocity model, $t$ is the time dimension and $s(t)$ is the source, i.e., a wavelet representing the seismic shot.
    
    The finite difference method (FDM) is often used to numerically solve \eqref{eq:ondaRTM} by approximating its PDEs (partial differential equations). Approximations of higher orders provide more accurate results, with smaller numerical errors. Spatial and time restrictions should be observed when solving finite differences by a numerical approach \cite{Carcione2002}.
    
    The content of the velocity model, $c(\mathbf{x})$, plays an important role from the geophysical perspective. Its complexity is the reason why RTM is used. It also influences the computational cost of the RTM as it determines the spatial and time resolutions. In other words, the maximum and minimum values of the velocity model, $c_{\text{min}}$ and $c_{\text{max}}$, directly influence on the total number of operations performed by an RTM. However, for a fixed $f_{\text{max}}$, two models having the same $c_{\text{min}}$ and $c_{\text{max}}$ demand the same time and spatial resolutions, and generally incurs in the same computational cost, no matter they have different geological structures.
    
    The wave propagation via FDM is performed over a limited grid representing the region of interest. Nevertheless, the region where the seismic survey takes place is not restricted to that region of interest. For this reason, it is common practice to add extra points to the limits of the grid, in order to absorb the energy reaching the borders of the model \cite{Cerjan1985}.
    
    RTM relies on the principle that the incident and the reflected wavefields, $u_\text{i}(\mathbf{x},t)$ and $u_\text{r}(\mathbf{x},t)$, correlate at the reflection interfaces. An image condition with the following mathematical description performs this correlation.
    \begin{equation} \label{eq:correlacao}
    I(\mathbf{x}) = \int\limits_{t=0}^{T} u_\text{i}(\mathbf{x},t) \cdot u_\text{r}(\mathbf{x},t)\text{d}t\text{,}
    \end{equation}
    where $T$ is the total time of the the simulation.
    
    The three RTM steps (as Algorithm \ref{alg:serialrtm} shows) are repeated for each shot location generating one migrated section per shot. The final migrated section is achieved by summing up all shot migrations.
    
    The RTM method used in the experiments described in this paper is an extension of the RTM introduced by Nunes-do-Rosario \textit{et al}. \cite{Nunes-do-rosario2015}. It is implemented in C with a hybrid parallel approach. MPI is used to distribute the workload of different shots among computational nodes of a distributed system, and OpenMP is employed to parallelize internal loops of each shot processing. The implemented parallel RTM code is described in Algorithm \ref{alg:rtm}.
    
    \begin{algorithm}
    \caption{Reverse time migration with work-stealing load balancing. $ns$ is the number of time steps. $i_\text{shot}$ is the number of the shot being processed.}
    \begin{algorithmic}[1]
    \label{alg:rtm}
    \STATE statically distribute shots among nodes using MPI \label{l:initws}
    \STATE read RTM parameters
    \STATE compute absorbing boundaries coefficients \label{l:absorb}
    \STATE \#OpenMP parallel section begin
    \FORALL{(shots locations of the process)} \label{l:slb}
        \STATE read shot seismogram
        \FOR {($t_i = 0$ to $ns$)} \label{l:flb}
            \STATE \#OpenMP for
            \FOR {(all grid points)}
                \STATE compute the wavefield \label{l:fwd}
            \ENDFOR
            \STATE add the source wavelet
            \STATE write wavefield to disk \label{l:writetodisk}
        \ENDFOR \label{l:fle}
        \FOR {($t_i = ns-1$ to $0$)} \label{l:blb}
            \STATE \#OpenMP for
            \FORALL {(grid points)}
                \STATE compute the wavefield \label{l:bwd}
            \ENDFOR
            \STATE \#OpenMP for
            \FORALL {(receivers location)}
                \STATE inject observed data samples at time $t_i$
            \ENDFOR
            \STATE read forward wavefield at $t_i$ from disk \label{l:readfromdisk}
            \STATE \#OpenMP for
            \FORALL {(main grid points)}
                \STATE perform image condition \label{l:imgcond}
            \ENDFOR
        \ENDFOR \label{l:ble}
    \ENDFOR \label{l:sle}
    \STATE \#OpenMP parallel section end
    \STATE reduce all nodes migrated sections \label{l:reduce}
    \end{algorithmic} 
    \end{algorithm}
    
    Absorbing boundaries are implemented in our RTM code as reduction coefficients (Line \ref{l:absorb} of Algorithm \ref{alg:rtm}) that taper the wavefield amplitudes in a layer of grid points surrounding the mesh as proposed in \cite{Cerjan1985}. The acoustic wave equation (\ref{eq:ondaRTM}) is solved for each propagation by the FDM with a second order approximation in time and eighth order approximation for each spatial dimension (Lines \ref{l:fwd} and \ref{l:bwd} of Algorithm \ref{alg:rtm}).
    
    The incident wavefield is stored on disk (Line \ref{l:writetodisk} of Algorithm \ref{alg:rtm}) at each forward wave propagation time step. At each backward wave propagation time step, the incident wavefield is read from disk (Line \ref{l:readfromdisk} of Algorithm \ref{alg:rtm}) in order to perform the image condition (Line \ref{l:imgcond} of Algorithm \ref{alg:rtm}).
    
    The load balancing of the RTM described by Algorithm \ref{alg:rtm} is static. An equal amount of shots, or nearly equal, is allocated to each node at the beginning of the algorithm (Line \ref{l:initws}). From this point on, no more load balancing decisions are taken. If a process finishes processing all its shots, i.e., leaves the shots loop (Lines from \ref{l:slb} to \ref{l:sle}), it has to wait for the slowest process in order to collectively summing up all shot migrations, i.e., performing the reduction operation, through the command MPI\_Reduce, in Line \ref{l:reduce}. Since MPI does not provide task schedulers, the static schedule is often used because of its ease of implementation.

\section{Cyclic Token-based Work-Stealing}
\label{sec:ctws}
    Cyclic token-based work-stealing (CTWS) is the load balancing method for distributed memory systems introduced in this paper. It is a library implemented in C using MPI. In order to reduce communication overhead, CTWS is implemented using MPI one-sided communication.
    
    Since MPI-2 \cite{Gropp1999}, MPI specification includes the concept of one-sided communication. This MPI feature implements RMA, which allows processes to make a portion of their local memory available for access by other processes. In one-sided communications, the process that accesses the memory is called the origin process while the process whose memory is accessed is called the target process \cite{Park2009}. All processes involved in one-sided communication must collectively create windows. A window is a structure with information on the memory regions which the processes make available for RMA.
    
    Many operations are available on MPI one-sided communication. Our work mainly uses the operations MPI\_Put and MPI\_Get, which are used to write to and read from remote memory, respectively. These operations are passive, i.e., the target process is not involved in the operation. Therefore, the target process keeps computing its tasks while the RMA operation is performed.
    
    A token and a list of remaining tasks per process are the two main elements of the proposed work-stealing technique. Both are implemented as MPI one-sided communication windows. The token was implemented as an integer number. It is initialized as $0$, meaning that, according to the list of remaining tasks, there are tasks to be stolen. The first process to figure out that no more tasks can be stolen sets the token to $1$, i.e., sets the token to finish. 
    
    The token gets passed around through an MPI\_Put operation in a round-robin fashion. Only the process owning the token can update the list of tasks per process and steal tasks. This strategy avoids deadlocks that would be caused by two processes trying to steal from each other at the same time. In such a case, both of them would have to grant access to both of their lists of remaining tasks. If both of them granted access to one of these lists, a deadlock would occur.
    
    In the initialization (Line \ref{l:initctws} of Algorithm \ref{alg:ctws}), the token must be allocated to a single process. The shots to be processed are equally distributed among the processes. Each process has its copy of this list of tasks per process. This list is implemented as an array of integers where the $i$-th element is the number of remaining tasks of the $i$-th process. The functions $getTask()$ and $updateList()$ are responsible for managing the token and the list of tasks.
    
    \begin{algorithm}
    \caption{Cyclic Token-based Work-Stealing. $t_{\text{id}}$ is the task identification number.}
    \begin{algorithmic}[1]
    \label{alg:ctws}
    \STATE Initialize CTWS variables \label{l:initctws}
    \STATE $t_{\text{id}} = getTask()$ \label{l:gettask1}
    \WHILE{($t_{id} \neq -1$)} \label{l:tlb}
        \FORALL{(iterations of task $t_{\text{id}}$)}
            \STATE $updateList()$ \label{l:chk}
            \STATE Compute an iteration of $t_{\text{id}}$
        \ENDFOR
        \STATE $t_{\text{id}} = getTask()$ \label{l:gettask2}
    \ENDWHILE \label{l:tle}
    \end{algorithmic} 
    \end{algorithm}
    
    The proposed strategy is designed for applications with iterative tasks. At each task iteration, the function $updateList()$ is called by each process. It first verify whether it possesses the token (Line \ref{l:chk} of Algorithm \ref{alg:ctws}). Should it have the token and it is not set to finish, the process updates its number of remaining tasks in its list and copies its list to the next process through an MPI\_Put operation in a ring fashion. When the process does not have the token, it simply continues working on its tasks. By doing so, any process has a close approximation of the current amount of remaining tasks of each process. This information is then used to lead the stealing stage.
    
\begin{figure}
    \centering
    \includegraphics[width=0.99\textwidth]{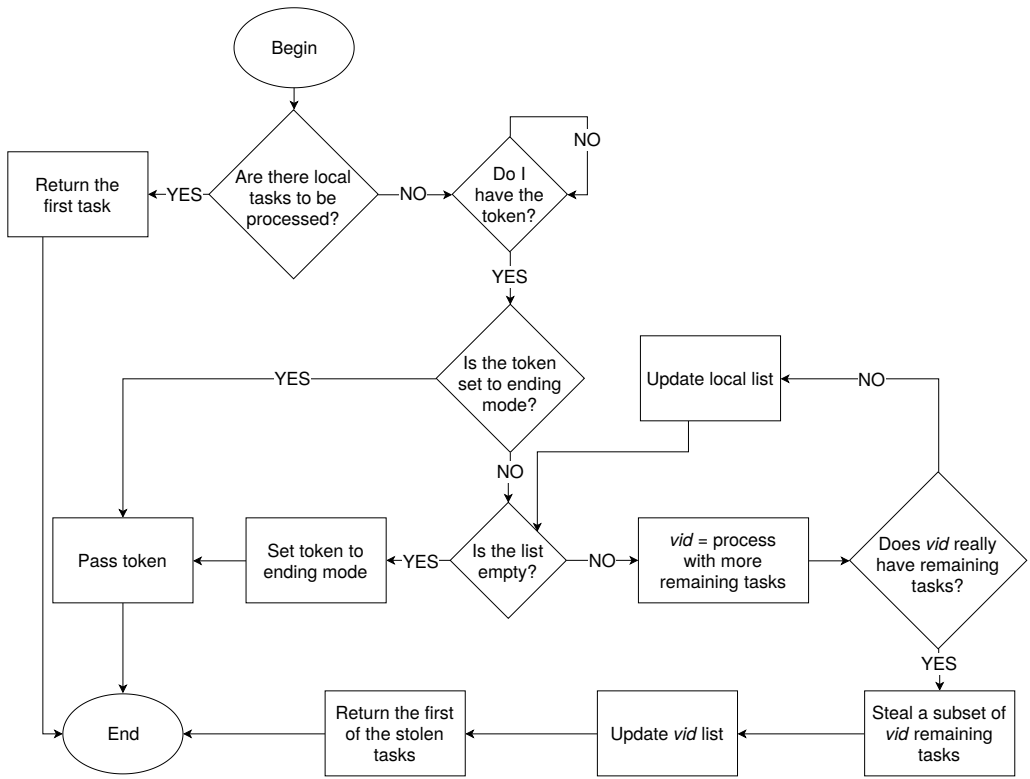}
    \caption{Detailed flow chart of the function $getTask()$ that is responsible for determining which is the next task to be processed by each process. In this work, the task unit to be processed is the RTM of each shot gather.}
    \label{fig:getTask}
    \end{figure}

    The core of the proposed work-stealing strategy is the function $getTask()$ (Lines \ref{l:gettask1} and \ref{l:gettask2} of Algorithm \ref{alg:ctws}), which is detailed by the flow chart of Fig. \ref{fig:getTask}. When a process has shots to be processed, $getTask()$ returns the first of them. Otherwise, the process attempts to steal tasks from other processes.
    
    Only the process possessing the token can try to steal tasks. For this reason, the first step of the proposed strategy is to ensure that the process has the token. If it does not, it will perform a busy-wait by repeatedly verifying whether it posses the token. Once the token arrives, the thief process checks whether the token is set to finish. If so, no work-stealing is needed, and the process continues to the reduction operation.
    
    However, when the token is not set to finish, the thief process tries to steal from the process with more remaining tasks, according to the thief's list of tasks. Since the list of remaining tasks per process is an approximation, the real number of remaining tasks may have changed by the stealing time. For this reason, the thief process verifies the actual number of remaining tasks of the victim process through an MPI\_Get operation. If the victim process does not have tasks to be stolen, the thief process updates its list of tasks and restarts the procedure by finding a new victim process in its updated list of tasks. Should there be no more tasks left to be stolen, then the thief process sets the token to finish, forwards it to the next process and continues to the reduction operation.
    
    When the thief process finds a victim process with tasks to be stolen, it uses MPI one-sided communication (MPI\_Put and MPI\_Get) to steal a subset of the remaining tasks of the victim process. For the tests in this work, half of the remaining tasks are stolen. As discussed by Dinan \textit{et al}. \cite{Dinan2009}, stealing half of the tasks of the victim increases the number of possible victims for the next steals. This strategy aims to improve scalability by reducing the time to locate and steal tasks. Finding an optimal number of tasks to be stolen is not of the scope of this paper.
    
    This stealing procedure is seamless to the victim process, i.e., the victim process will not stop processing its current task to communicate with the thief process. At this point, the thief process starts to work on the first of its stolen tasks. Should the victim process have a single remaining task, then the thief process will try to steal it. In case the victim process also tries to start processing its only left task at the same time, the race condition is avoided by the one-sided communication operators $MPI\_Win\_lock$ and $MPI\_Win\_unlock$ set to the type $MPI\_LOCK\_EXCLUSIVE$. These commands ensure mutual exclusion allowing a single process to access the window at a time.
    
\section{CTWS Applied to RTM}
\label{sec:wsrtm}
    In this work, the task unit to be processed is the RTM of each shot gather. In other words, the iterations of the shots loop (Lines from \ref{l:slb} to \ref{l:sle} of Algorithm \ref{alg:rtm}) are distributed to the nodes of a distributed system using CTWS. For this reason, the commands controlling the shots loop of the RTM must be replaced by the commands controlling the tasks loop of CTWS. Line \ref{l:slb} of Algorithm \ref{alg:rtm} is replaced by Lines \ref{l:gettask1} and  \ref{l:tlb} of Algorithm \ref{alg:ctws}, and Line \ref{l:sle} of Algorithm \ref{alg:rtm} is replaced  by Lines \ref{l:gettask2} and \ref{l:tle} of Algorithm \ref{alg:ctws}. In this context, the task identification number, $t_{\text{id}}$, represents the number of the shot gather. The function $updateList()$ is called inside of both the forward propagation loop (Lines from \ref{l:flb} to \ref{l:fle} of Algorithm \ref{alg:rtm}) and the backward propagation loop (Lines from \ref{l:blb} to \ref{l:ble} of Algorithm \ref{alg:rtm}).
    
    The larger the number of processes, the shorter the time that each process will have the token. Because of that, the overhead caused by $updateList()$ in the RTM is proportionally smaller for larger numbers of processes and larger input sizes. On the other hand, by running $updateList()$ and having the token in each process fewer times, the list of remaining tasks per process is more prone to be out of date. This way the number of unsuccessful steals attempts performed by $getTask()$ may increase and so its overhead.

\section{Results and Discussion}
\label{sec:results}
    The experiments were performed on Yemoja, an $856$ node supercomputer. Each computational node hosts two processors $10$-core Intel Xeon E5-2690 Ivy Bridge v2 at $3.00~\text{GHz}$. $200$ nodes are equipped with $256~\text{GB}$ RAM and $656$ nodes with $128~\text{GB}$ RAM. This supercomputer employs an $850$ TB Lustre parallel distributed file system. Yemoja is located at the Manufacturing and Technology Integrated Campus of the National Service of Industrial Training (SENAI-CIMATEC). Both the $128~\text{GB}$ RAM and the $256~\text{GB}$ RAM were used in the following experiments. Since the total amount of RAM required by our RTM implementation is significantly inferior to $128~\text{GB}$, this fact does not influence the algorithm performance.
    
    In order to validate the 3D wave propagator, which underlies the 3D RTM algorithm used in the experiments, we compared a seismic trace generated by our propagator with the analytical solution, computed according to \cite{Hoop1960}, in a homogeneous velocity model. The source was a Ricker wavelet with a peak frequency of $20~\text{Hz}$. The distance between source and receiver is $200~\text{m}$. The medium has a constant velocity of $2000~\text{m/s}$. In this experiment, our wave propagator provided a very accurate approximation to the 3D waveform analytical solution with a mean squared error of $6\times10^{-14}$.
    
    For the following experiments, the size of the input grid is $401\times 401 \times 401$, the peak frequency of the source wavelet is $20~\text{Hz}$, the time sampling is $1~\text{ms}$, the spatial sampling is $10~\text{m}$, and the number of time steps is $3501$.  $c(\mathbf{x})$ is a two layers model with a horizontal interface positioned at the center of the vertical dimension. The velocity is $1400~\text{m/s}$ for the top layer and $2000~\text{m/s}$ for the bottom layer.
    
    The programs were compiled with the \textit{gcc} compiler using the optimization flag \textit{-O3} and OpenMPI $3.1.2$ for all experiments. A single MPI process was created at each computational node. We used HPCToolkit performance tools \cite{Adhianto2010} to measure the execution times and the overhead of our strategy. For all the following experiments using CTWS, the load balancing overhead was inferior to $0.4\%$. A single experiment was performed at a time in order to avoid multiple tests competing for the shared resources of the cluster.
    
\begin{figure}
    \centering
    \includegraphics[width=0.99\textwidth]{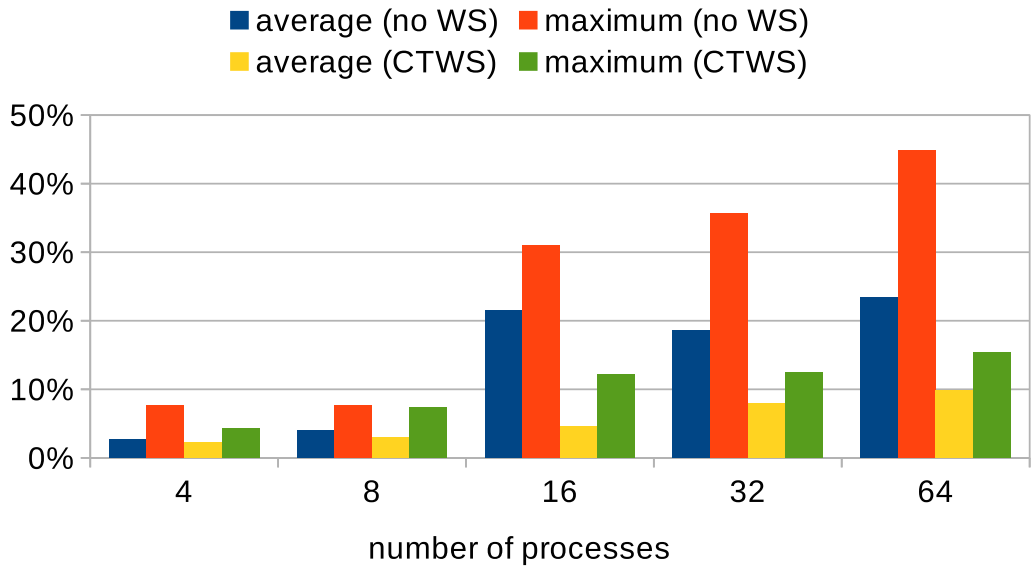}
    \caption{3D RTM maximum process idle time and average process idle time with $4$, $8$, $16$, $32$ and $64$ nodes. Both RTM implementations with and without the proposed work-stealing method (CTWS) process 10 shots per node.}
    \label{fig:idletime}
    \end{figure}

\begin{figure}
    \centering
    \includegraphics[width=0.99\textwidth]{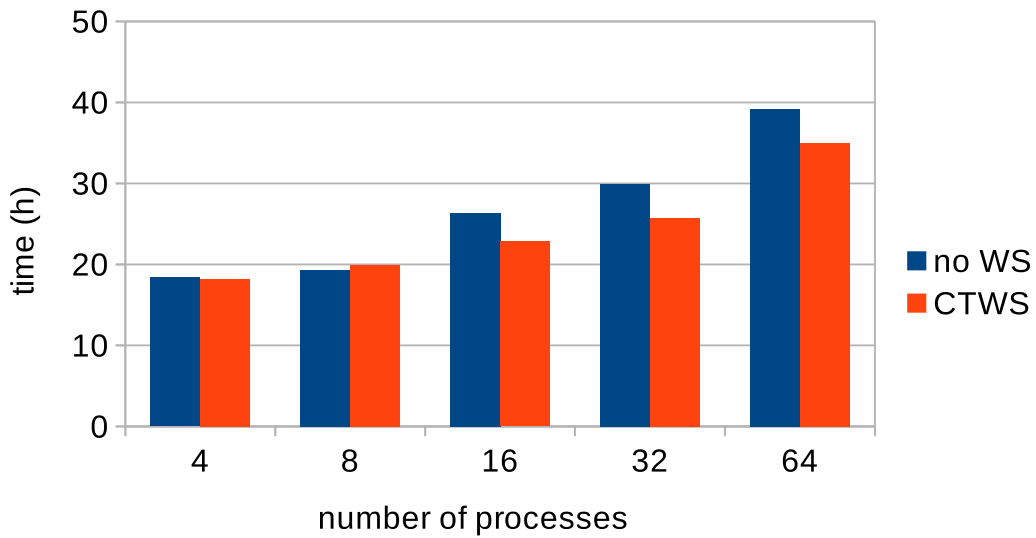}
    \caption{3D RTM total runtime with $4$, $8$, $16$, $32$ and $64$ nodes. Both RTM implementations with and without the proposed work-stealing method (CTWS) process 10 shots per node.}
    \label{fig:runtime}
    \end{figure}

    Firstly, we measured the load imbalance of the 3D RTM without applying a dynamic load balancing technique. For that we ran the RTM of $40,\, 80,\, 160,\, 320\, \text{and}\, 640$ shots with $4,\, 8,\, 16,\, 32\, \text{and}\, 64$ nodes, respectively. As shown in Fig. \ref{fig:idletime} and \ref{fig:runtime}, for the experiment with $4$ nodes, the average idle time per node is $2.7\,\%$ of the total time of $18.4\,\text{h}$. As the number of nodes increases up to $64$, the average idle time per node increases to $23.4\,\%$ of the total time of $39.2\,\text{h}$. Although the number of shots per node is the same for each experiment, the competition for shared resources of the cluster (e.g., network and parallel file system) increases the runtime as the number of nodes increase.
    
\begin{figure}
    \centering
    \includegraphics[width=0.99\textwidth]{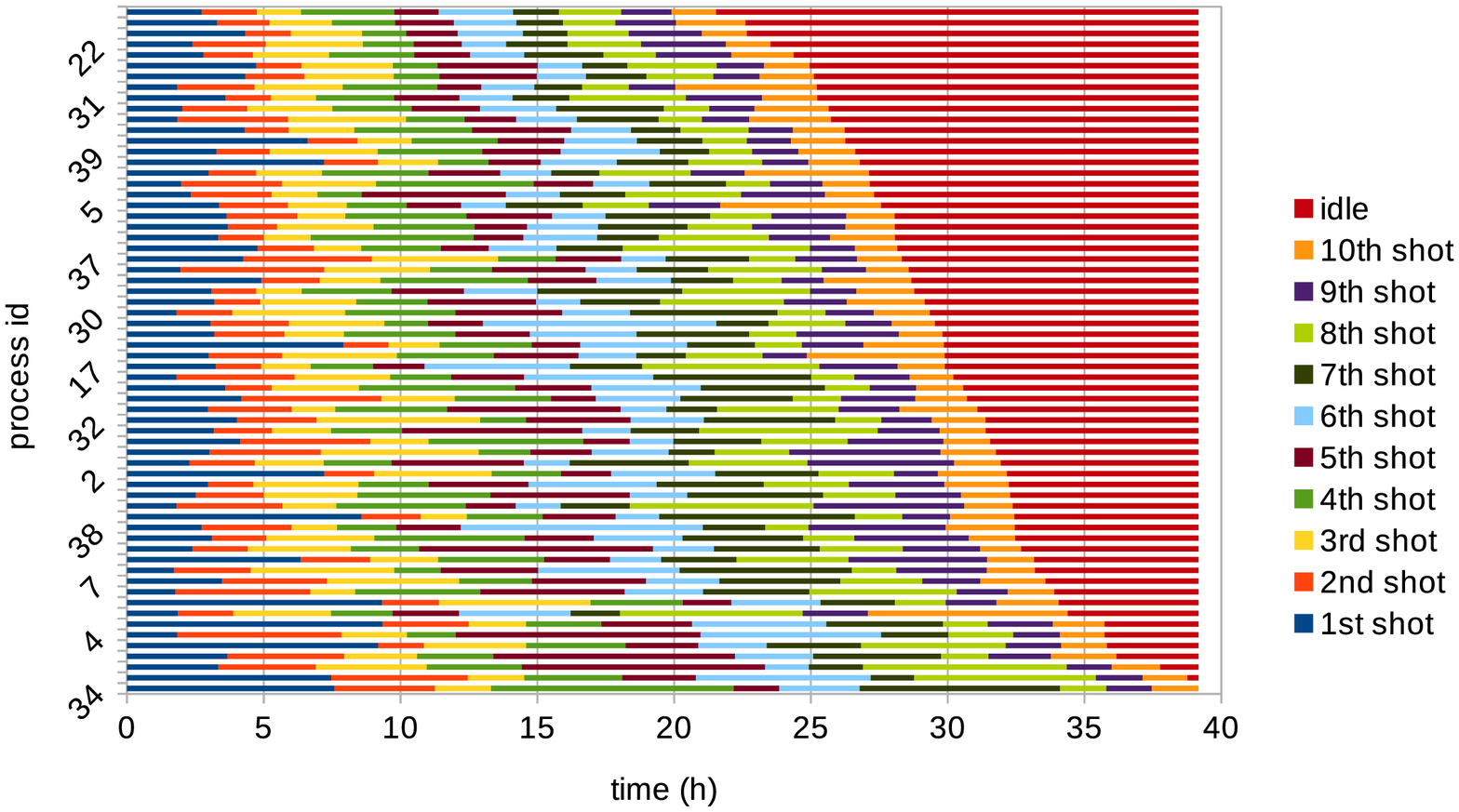}
    \caption{Example of 3D RTM runtime per process and shot ran over $64$ nodes. The shots are numbered in the order they are processed in the node they were assigned to. The processes are sorted by their idle time.}
    \label{fig:DRTM64-Yemoja}
    \end{figure}

    Fig. \ref{fig:DRTM64-Yemoja} details the execution of the 3D RTM ran over $64$ nodes without applying a dynamic load balancing technique. Although the workload is distributed evenly among the homogeneous nodes, the runtime of a single shot RTM ranges from $1.5$ to $9.3~\text{h}$. The fastest node stays idle for $17.6~\text{h}$ while the other nodes finish their tasks, i.e., $45\%$ of the total runtime. Factors as a race condition for the network and the parallel storage system can cause such load imbalance.
      
    Fig. \ref{fig:runtime} also shows results generated by the proposed work-stealing technique employed in the same set of experiments. The proposed technique presented a maximum average idle time of $9.9\%$, showing its effectiveness in balancing the load. For the experiment with $4$ nodes, the average idle time per node is $2.3\,\%$ of the total time of $18.2\,\text{h}$. As the number of nodes increases up to $64$, the average idle time per node slightly increases to $9.9\,\%$ of the total time of $34.9\,\text{h}$.
    
    The total execution times displayed in Fig. \ref{fig:runtime} show that the proposed technique outperforms the 3D RTM with the conventional static load balancing when using a more substantial number of nodes. The total runtime was reduced by $13.4\%$, $14.1\%$ and $10.8\%$ when ran over $16$, $32$ and $64$ nodes, respectively. For the fewest number of nodes, however, the proposed technique performance was similar to the static load balancing. The total runtime increased by $3.2\%$ when ran over $8$ nodes and decreased by $0.9\%$ when ran over $4$ nodes. Industry-scale RTM, however, is usually performed over a large number of computational nodes. Regarding the load imbalance, as shown in Fig. \ref{fig:idletime}, the proposed method was able to reduce the average idle time for all the performed tests. In the best scenario, when ran over $16$ nodes, the idle time was reduced from $21.5\,\%$ to $4.5\,\%$, representing a $78.4\,\%$ improvement in the effective use of the resources.
    
\begin{figure}
    \centering
    \includegraphics[width=0.99\textwidth]{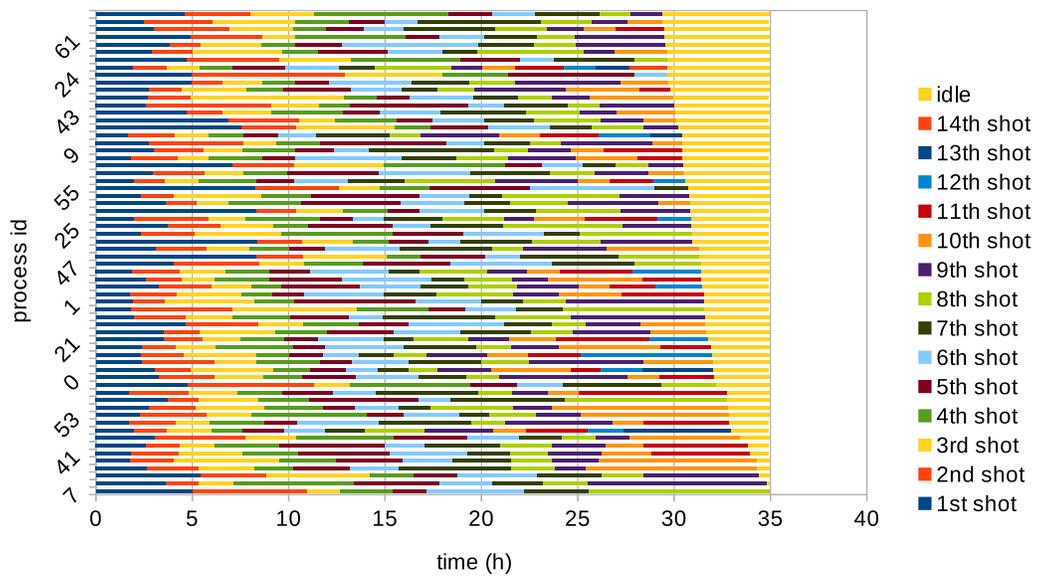}
    \caption{Example of 3D RTM runtime per process and shot ran over $64$ nodes using the proposed work-stealing. The shots numbers refer to the order they were processed within its node. The processes are sorted by their idle time.}
    \label{fig:WSDRTM64-Yemoja}
    \end{figure}

    Fig. \ref{fig:WSDRTM64-Yemoja} details the 3D RTM execution over 64 nodes, using the proposed work-stealing technique. Although there are differences between the processing times of a single shot, nodes with better resource availability steal shots from others that are slower, thus improving load balancing. The least busy node processed only $6$ shots while the busiest node processed $14$ shots. 
      
    Table \ref{tab:steals64} presents all steal attempts of the example of Fig. \ref{fig:WSDRTM64-Yemoja}. It shows that $37$ out of $38$ steal attempts were successful. In five cases, $13.2~\%$ of the steals, a task was stolen for the second time:
    \begin{enumerate}
        \item in steal 18, process 20 stole the task 387 from process 21. Before that, in stealing 5, process 21 stole the same task from process 38;
        \item in steal 23, process 28 stole the task 178 from process 29. Before that, in stealing 8, process 29 stole the same task from process 17;
        \item in steal 24, process 34 stole the task 478 from process 9. Before that, in stealing 10, process 9 stole the same task from process 47;
        \item in steal 28, process 21 stole the task 307 from process 16. Before that, in stealing 3, process 16 stole the same task from process 30;
        \item in steal 29, process 41 stole the task 318 from process 18. Before that, in stealing 9, process 18 stole the same task from process 31;
    \end{enumerate}
    
    In our test case, stealing the same task multiple times does not represent an additional overhead since there is no significant extra cost to move a task between processes. This fact occurs because all the data is available to all processes through the Yemoja's parallel file system. A method that considers the cost of moving tasks is left to future work.
    
    \begin{table}
    \caption{Steal attempts of the example of Fig. \ref{fig:WSDRTM64-Yemoja}.}
    \centering
    \begin{tabular}{|c|c|c|c|}
    \hline
    \textbf{stealing attempt} & \textbf{thief process} & \textbf{victim process} & \textbf{stolen tasks} \\ \hline
    1 & 63 & 20 & 205,206,207 \\ \hline
    2 & 19 & 24 & 245,246,247 \\ \hline
    3 & 16 & 30 & 305,306,307 \\ \hline
    4 & 49 & 34 & 346,347 \\ \hline
    5 & 21 & 38 & 386,387 \\ \hline
    6 & 10 & 55 & 556,557 \\ \hline
    7 & 22 & 7 & 77,78 \\ \hline
    8 & 29 & 17 & 177,178 \\ \hline
    9 & 18 & 31 & 317,318 \\ \hline
    10 & 9 & 47 & 477,478 \\ \hline
    11 & 15 & 8 & 88 \\ \hline
    12 & 50 & 51 & 518 \\ \hline
    13 & 30 & 24 & 248 \\ \hline
    14 & 62 & 25 & 258 \\ \hline
    15 & 32 & 55 & 558 \\ \hline
    16 & 63 & 57 & 578 \\ \hline
    17 & 17 & 59 & 598 \\ \hline
    18 & 20 & 21 & 387 \\ \hline
    19 & 2 & 1 & 19 \\ \hline
    20 & 23 & 4 & 49 \\ \hline
    21 & 22 & 5 & 59 \\ \hline
    22 & 42 & 8 & 89 \\ \hline
    23 & 28 & 29 & 178 \\ \hline
    24 & 34 & 9 & 478 \\ \hline
    25 & 35 & 12 & - \\ \hline
    26 & 35 & 13 & 139 \\ \hline
    27 & 30 & 14 & 149 \\ \hline
    28 & 21 & 16 & 307 \\ \hline
    29 & 41 & 18 & 318 \\ \hline
    30 & 49 & 37 & 379 \\ \hline
    31 & 13 & 38 & 389 \\ \hline
    32 & 15 & 39 & 399 \\ \hline
    33 & 50 & 54 & 549 \\ \hline
    34 & 18 & 57 & 579 \\ \hline
    35 & 5 & 58 & 589 \\ \hline
    36 & 0 & 59 & 599 \\ \hline
    37 & 3 & 60 & 609 \\ \hline
    38 & 38 & 61 & 619 \\ \hline
    \end{tabular}
    \label{tab:steals64}
    \end{table}
    
    Table \ref{tab:steals} shows the total number of steal attempts and failed steals varying the number of processes. In this test set, $1$ out of $55$ steal attempts was unsuccessful, i.e., $98.2\%$ of the steal attempts were successful.
    
    \begin{table}
    \caption{Steal attempts varying the number of processes.}
    \centering
    \begin{tabular}{|c|c|c|c|c|c|}
    \hline
    \textbf{number of processes} & 4 & 8 & 16 & 32 & 64 \\ \hline
    \textbf{total steal attempts} & 0 & 2 & 3 & 12 & 38 \\ \hline
    \textbf{failed steal attempts} & 0 & 0 & 0 & 0 & 1 \\ \hline
    \end{tabular}
    \label{tab:steals}
    \end{table}
    
    In terms of weak scalability, both the RTM with and without CTWS are not scalable as the total runtime increases when the number of shots and nodes are doubled. This means that RTM with CTWS may also be affected by the concurrency for shared resources of the distributed system as the number of nodes increases. However, by moving tasks to nodes with better resource availability, RTM with CTWS was able to deliver up to $14.1\%$ speedup when compared to using a static load distribution.
    
\section{Related Works}
\label{sec:literature}
    Several authors have proposed strategies to address the load imbalance for shared memory systems. Barros \textit{et al}. \cite{Barros2018} introduced a runtime method based on coupled simulated annealing (CSA) \cite{Xavier-de-Souza2010} to auto-tune the workload distribution of 3D acoustic wave propagation implemented with the FDM method. Andreolli \textit{et al}. \cite{Andreolli2014,Andreolli2015} proposed a compilation-time auto-tuning based on genetic algorithms to find the best set of parameters (e.g., workload distribution, compilation flags) for seismic applications. Sena \textit{et al}. \cite{Sena2011} used cache blocking for the 3D RTM and proposed a procedure called Min-Worst-Min Block (MWMB) to find an efficient block size. Hofmeyr \textit{et al}. \cite{Hofmeyr2011} introduced a dynamic load balancing for multicore systems using runtime tools. Tchiboukdjian \textit{et al}. \cite{Tchiboukdjian2011} proposed a method that ensures all the data in the cache memory is used before being replaced. This method was designed for applications with linear access to memory. Imam and Sarkar \cite{Imam2015} presented a work-stealing scheduler based on task priority queues. Balancing the computational load at the shared memory level can lead to a significant reduction in the execution time. Our work aims to achieve further improvement by balancing the workload at the distributed memory level.
    
    Other authors provide methods to deal with the load imbalance of distributed memory systems. Khaitan \textit{et al}. \cite{Khaitan2013} proposed a master-slave based load balancing approach. Tesser \textit{et al}. \cite{Tesser2014,Tesser2017,Tesser2018} proposed a simulation-based strategy to evaluate the performance and tune the dynamic load balancing of iterative MPI applications and applied it to a 3D wave propagation. Padoin \textit{et al}. \cite{Padoin2014,Padoin2017} proposed combining a load balancing with techniques of processor frequency control in order to reduce energy consumption along with execution time. These approaches differ from this work for being centralized, i.e., a single or a few computational processes take the load balancing decisions. This behavior may lead to overload at the central element and significantly degrade performance \cite{Khaitan2013}.
    
    To avoid losses of performance caused by a centralized load balancing element, some authors proposed decentralized load balancing strategies. Sharma and Kanungo \cite{Sharma2014} presented a technique to balance the computational load in heterogeneous multicore clusters, where no prior knowledge about the computational resources is required. Zheng \textit{et al}. \cite{Zheng2011} introduced a periodic load balancing strategy, where the balancing decisions are taken hierarchically in a tree fashion. Different from this work, in these methods, the processes involved in the load balancing decisions have to synchronize to exchange information. This communication synchronization overhead may reduce parallel performance.
    
    Work-stealing algorithms \cite{Blumofe1999} have been used to provide decentralized load balancing methods for distributed systems. Martinez \textit{et al}. \cite{Martinez2016} used StarPU, a task-based runtime system, to distribute the load balance of the 3D isotropic elastic wave propagation among processors and graphics processing units (GPUs) simultaneously. They compared centralized load-balancing and decentralized work-stealing algorithms from StarPU. Khaitan and Mccalley \cite{Khaitan2014} applied dynamic load balancing with work-stealing to a contingency analysis application while Mor and Maillard \cite{Mor2011} proposed an MPI library for load balancing branch and bound applications. These approaches use asynchronous communication to reduce the communication overhead as the processes which originate the communication may keep working while waiting for replies from their messages. Different from our work, these papers employ two-sided communication, i.e., both the origin and the destination processes are involved in the communication. This way, the destination processes have to interrupt their computation eventually to reply to the messages they have received. Also, this kind of non-blocking communication may imply in the origin process having to wait for its reply even when overlapping communication with computation.

    One-sided communication is an alternative to reduce communication overhead. This model of communication allows a process to read and write data from a remote memory region without the target process being involved. Some authors have used it in the recent literature. Li \textit{et al}. \cite{Li2013} used profiling information to estimate the task grain size and guide the asynchronous work-stealing. Kumar \textit{et al}. \cite{Kumar2016} introduced a load-aware work-stealing based on a policy to choose a victim that completely avoids the failed steals. Dinan \textit{et al}. \cite{Dinan2009} discussed the design and scalability aspects of work-stealing for distributed memory systems. They also proposed a runtime system for supporting work-stealing, which implements several techniques to achieve scalability in distributed memory systems. These methods employ one-sided communication through partitioned global address space (PGAS), a programming model that provides a globally shared address space for distributed memory. On the contrary, our work employs relies on MPI one-sided communication, which has no global address space. According to Fu \textit{et al}. \cite{Fu2018}, employing PGAS often demands an important development effort to exploit these programming models thoroughly. Moreover, a vast majority of scientific codes use MPI either directly or via third-party software.

    Other authors have recently used MPI implementations of one-sided communication in areas such as large-scale multimedia content analysis \cite{Essafi2017}, graph processing \cite{Fu2018} and matrix operations \cite{Ghosh2016,Lazzaro2017}. According to Diaz \textit{et al}. \cite{Diaz2012}, MPI has been the \textit{de facto} standard in HPC for the last decades. In the context of load balancing, Vishnu and Agarwal \cite{Vishnu2015} introduced a work-stealing method using MPI one-sided communication for machine learning and data mining algorithms. Different from our proposal, their approach for victim selection is either random, which may increase the number of network requests, or prone to network contention because of having multiple thief processes trying to steal the same victim. Moreover, Vishnu and Agarwal employ a termination strategy that does not look at the entire victim set, potentially causing some processes to finish while there are remaining tasks to perform. Differently, we employ a termination strategy in which the processes only finish when there are no more victims to be stolen, and the token is set to finish. This is only achieved because, in our proposed work-stealing algorithm, we share and update \emph{global load information} without the need for synchronization. Sharing and updating the global load information has the main advantages of helping the thief processes to make better decisions and preventing failed stealing attempts.
    
    Regarding load balancing strategies to RTM, little effort has been employed to schedule RTM tasks among nodes of distributed systems. Several authors implement parallel RTM for distributed systems using static scheduling \cite{Abdelkhalek2012,Qawasmeh2017,Akanksha2017,Chu2008,Perrone2012,Chu2009,Lu2013,Paul2016}. As shown in Section \ref{sec:results}, the use of static distribution may lead to inefficient use of the distributed computational resources as faster nodes may wait idly for the slower ones to finish their tasks. On the other hand, our work-stealing strategy allows moving tasks among nodes in order to keep all resources busy as much as possible. In this work, we compare our proposed load balancing approach to static scheduling as it is arguably the conventional strategy to distribute RTM shots among the computing nodes in distributed systems.
    
    In summary, this work distinguishes itself from the others as it proposes a decentralized work-stealing method to balance the load of the RTM in distributed systems. It employs MPI one-sided communication to reduce its overhead by communicating asynchronously and without the victim's involvement. By keeping global load information, our method lowers the cost of victim selection and process termination. Consequently, it reduces the number of failed stealing attempts.
    
    Table \ref{tab:literature} presents a summary of the works related to load balancing mentioned above, highlighting their main characteristics in comparison to the method proposed in this work.
    
    \begin{table*}
    \caption{Literature review on load balancing methods. The proposed work-stealing method benefits from MPI one-sided communication to further reduce communication overhead and is applied to RTM in distributed memory systems.}
    \centering
    \footnotesize
    \hspace*{-3.2cm}\begin{tabular}{|c|c|c|c|c|c|c|c|c|c|}
    \hline
    ~ & ~ & ~ & ~ & asynchronous & one-sided & ~ & ~ & WS with & ~\\
    ~ & distributed & decen- & work- & commu- & commu- & PGAS & MPI & global load & RTM\\
    ~ & memory & tralized & stealing & nication & nication & ~ & RMA & information & ~\\ \hline
    Barros \textit{et al}. \cite{Barros2018} & ~ & ~ & ~ & ~ & ~ & ~ & ~ & ~ & ~\\ \hline
    Andreolli \textit{et al}. \cite{Andreolli2014} & ~ & ~ & ~ & ~ & ~ & ~ & ~ & ~ & ~\\ \hline
    Andreolli \textit{et al}. \cite{Andreolli2015} & ~ & ~ & ~ & ~ & ~ & ~ & ~ & ~ & ~\\ \hline
    Sena \textit{et al}. \cite{Sena2011} & ~ & ~ & ~ & ~ & ~ & ~ & ~ & ~ & x\\ \hline
    Hofmeyr \textit{et al}. \cite{Hofmeyr2011} & ~ & x & ~ & ~ & ~ & ~ & ~ & ~ & ~\\ \hline
    Tchiboukdjian \textit{et al}. \cite{Tchiboukdjian2011} & ~ & ~ & x & ~ & ~ & ~ & ~ & ~& ~\\ \hline
    Imam and Sarkar \cite{Imam2015} & ~ & x & x & ~ & ~ & ~ & ~ & x & ~\\ \hline
    Khaitan \textit{et al}. \cite{Khaitan2013} & x & ~ & ~ & ~ & ~ & ~ & ~ &  & ~\\ \hline
    Tesser \textit{et al}. \cite{Tesser2014} & x & ~ & ~ & ~ & ~ & ~ & ~ & ~ & ~\\ \hline
    Tesser \textit{et al}. \cite{Tesser2017} & x & ~ & ~ & ~ & ~ & ~ & ~ & ~ & ~\\ \hline
    Tesser \textit{et al}. \cite{Tesser2018} & x & ~ & ~ & ~ & ~ & ~ & ~ & ~ & ~\\ \hline
    Padoin \textit{et al}. \cite{Padoin2014} & x & ~ & ~ & ~ & ~ & ~ & ~ & ~ & ~\\ \hline
    Padoin \textit{et al}. \cite{Padoin2017} & x & ~ & ~ & ~ & ~ & ~ & ~ & ~ & ~\\ \hline
    Sharma and Kanungo \cite{Sharma2014} & x & x & ~ & ~ & ~ & ~ & ~ & x & ~ \\ \hline
    Zheng \textit{et al}. \cite{Zheng2011} & x & x & ~ & ~ & ~ & x & ~ & ~ & ~\\ \hline
    Martinez \textit{et al}. \cite{Martinez2016} & x & x & x & x & ~ & ~ & ~ & ~ & ~\\ \hline
    Khaitan and Mccalley \cite{Khaitan2014} & x & x & x & x & ~ & ~ & ~ &  & ~\\ \hline
    Mor and Maillard \cite{Mor2011} & x & x & x & x & ~ & ~ & ~ &  & ~\\ \hline
    Li \textit{et al}. \cite{Li2013} & x & x & x & x & x & x & ~ & ~ & ~\\ \hline
    Kumar \textit{et al}. \cite{Kumar2016} & x & x & x & x & x & x & ~ & ~& ~ \\ \hline
    Dinan \textit{et al}. \cite{Dinan2009} & x & x & x & x & x & x & ~ & ~ & ~\\ \hline
    Vishnu and Agarwal \cite{Vishnu2015} & x & x & x & x & x & ~ & x & ~ & ~\\ \hline
    Our proposal & x & x & x & x & x & ~ & x & x & x\\ \hline
    \end{tabular}
    \label{tab:literature}
    \end{table*}
    
\section{Conclusions}
\label{sec:conclusions}
    We have presented a decentralized work-stealing strategy with asynchronous communication to balance the load of a 3D reverse time migration for distributed computing systems. Each process communicates in a round-robin fashion to maintain a close approximation of the remaining tasks list. This list is used to lead the stealing when processes are idle. This strategy decentralizes the dynamic load balancing and avoids the overhead of centralized decisions. A token avoids deadlocks by ensuring that two processes cannot steal each other at the same time. The MPI one-sided communication prevents race conditions by serializing access to a memory space by multiple processes. By using MPI one-sided communication, the stealing is seamless to the victim processes since they do not stop processing their tasks during the stealing, avoiding unnecessary communication.
    
    In the presented experiments, the 3D RTM had up to $23.4\,\%$ of average idle time when ran over $64$ nodes. This imbalance might be significantly reduced should the proposed work-stealing be applied. For the set of experiments performed in this paper, the proposed method has reduced the total execution time of the 3D RTM in up to $14.1\,\%$ and its load imbalance in the order of $78.4\,\%$ when compared to the conventional static distribution.
    
    Further investigation is necessary to assess whether additional improvement can be achieved by adjusting the frequency of checking the token, the number of shots to be stolen, the method used to update the list of remaining tasks and the technique to avoid deadlocks. Also, future work should focus on different aspects of a distributed system that may influence the load imbalance such as the use of fault tolerance protocols (e.g., \cite{Shang2018a,Shang2018b}) and heterogeneous computational systems. A comparison of our method against other load balancing methods is left to future work.

\section*{Acknowledgment}
    The authors gratefully acknowledge support from Shell Brazil through the project ``\textit{Novos M\'etodos de Explora\c{c}\~ao S\'ismica por Invers\~ao Completa das Formas de Onda}'' at the Universidade Federal do Rio Grande do Norte (UFRN) and the strategic importance of the support given by ANP through the R\&D levy regulation. 
    The authors are also thankful to CNPq (\textit{Conselho Nacional de Desenvolvimento Cient\'{i}fico e Tecnol\'{o}gico}) and CAPES (\textit{Coordena\c{c}\~{a}o de Aperfei\c{c}oamento de Pessoal de N\'{i}vel Superior}) for partially funding this research and to the High-Performance Computing Center at UFRN (NPAD/UFRN) and the Manufacturing and Technology Integrated Campus of the National Service of Industrial Training (SENAI CIMATEC) for making computer resources available.
    Finally, the authors thank Jorge Lopez from Shell and the anonymous reviewers for providing essential comments on this paper.

\bibliographystyle{IEEEtran}
\bibliography{ws}

%\EOD

\end{document}